# EDIS: A Simulation Software for Dynamic Ion Intercalation/Deintercalation Processes in Electrode Materials

Liqi Wang[a, b], Ruijuan Xiao[a, b, *] and Hong Li[a, b]

a Institute of Physics, Chinese Academy of Sciences, Beijing, 100190, China.

b School of Physical Sciences, University of Chinese Academy of Sciences, Beijing 100049, China

*Corresponding author. Email: rjxiao@iphy.ac.cn

**Abstract**

As the core determinant of lithium-ion battery performance, electrode materials play a crucial role in defining the battery's capacity, cycling stability, and durability. During charging and discharging, electrode materials undergo complex ion intercalation and deintercalation processes, accompanied by defect formation and structural evolution. However, the microscopic mechanisms underlying processes such as cation disordering, lattice oxygen loss, and stage structure formation phenomena are still not fully understood. To address these challenges, we have developed the Electrode Dynamic Ion Intercalation/Deintercalation Simulator (EDIS), a software platform designed to simulate the dynamic processes of ion intercalation and deintercalation in electrode materials. Leveraging high-precision machine learning potentials, EDIS can efficiently model structural evolution and lithium-ion diffusion behavior under various states of charge and discharge, achieving accuracy approaching that of quantum mechanical methods in relevant chemical spaces. The software supports quantitative analysis of how variations in lithium-ion concentration and distribution affect lithium-ion transport properties, enables evaluation of the impact of structural defects, and allows for tracking of both structural evolution and transport characteristics during continuous cycling. EDIS is versatile and can be extended to sodium-ion batteries and related systems. By enabling in-depth analysis of these microscopic processes, EDIS provides a robust theoretical tool for mechanistic studies and the rational design of high-performance electrode materials for next-generation lithium-ion batteries.

# 1. Introduction

As the core component of modern electrochemical energy storage systems, the performance of electrode materials directly determines key metrics of lithium-ion batteries, such as energy density, power characteristics, and cycle life.[1-2] With the rapid development of new energy vehicles and large-scale energy storage technologies, the demand for high-performance electrode materials is steadily increasing.[3-5] However, during charging and discharging, electrode materials undergo complex multiscale dynamic evolution.[6-15] At the atomic scale, the crystal lattice of active materials experiences ion intercalation and deintercalation, resulting in changes to lattice parameters and electronic structure.[6-7] At the mesoscale, materials may develop cracks, undergo phase transitions, or experience other forms of structural reconstruction.[8-12] At the macroscopic scale, these microscopic changes ultimately manifest as capacity fade and increased impedance in batteries.[13-15] A deep understanding of the interrelationships between these multiscale dynamic processes is essential for improving the performance of electrode materials.

Traditional experimental approaches can reveal macroscopic properties, but they often hard to capture atomic-scale dynamic details.[16-17] Theoretical simulation methods serve as powerful complements to experiments, but current mainstream simulation tools face limitations in accuracy, efficiency, and the targeted analysis of dynamic processes[18-19]. For example, first-principles methods offer high accuracy but are computationally expensive and cannot address long time-scale processes under practical conditions[20-23]. Classical molecular dynamics is efficient but limited by the accuracy of empirical force fields[24]. In addition, existing software lacks specialized analysis capabilities for ion transport and dynamic structural evolution in electrode materials.

To address these challenges, we have developed the Electrode Dynamic Ion Intercalation/Deintercalation Simulator (EDIS). EDIS leverages machine learning potential technology, combining the accuracy of first-principles calculations with the efficiency of molecular dynamics to enable precise and efficient simulation as well as in-depth analysis of the dynamic charging/discharging processes in electrode materials.[25-29] EDIS supports quantitative evaluation of ion transport properties, identification of intercalation and deintercalation sites, and tracking of structural evolution, providing a systematic tool for elucidating microscopic mechanisms and structure-property relationships in materials. Technically, EDIS is developed primarily in Python and integrates mainstream molecular dynamics platforms such as Large-scale Atomic/Molecular Massively Parallel Simulator (LAMMPS).[30] With hardware support including GPU acceleration, EDIS enables efficient simulations of large-scale systems and long-time scales. The simulation output data can be further analyzed using visualization tools such as Open Visualization Tool (OVITO).[31] At present, EDIS has been successfully applied to the study of electrode materials such as $LiNiO_2$ cathode, enabling the tracking of ion transport behavior and structural evolution during continuous charging/discharging process. Notably, with simple parameter adjustments, EDIS can also be extended to emerging energy storage systems such as sodium-ion batteries. This software will offer new tools and perspectives for researchers in related fields and promote further advancement in the study of electrode materials, providing important support for understanding capacity fading mechanisms and material optimization.

## 2. Theoretical foundations

Electrode materials undergo complex physicochemical changes during the charge/discharge cycles of lithium ion batteries. These changes significantly influence their electrochemical performance and cycle stability. In this chapter, we discuss the structural evolution, underlying mechanisms, and their impacts on material performance during cycling, using $LiNiO_2$ as a representative cathode material and graphite as a typical anode material. After clarifying the structural and property changes occurring in these materials during operation, this chapter further outlines computational strategies that can effectively probe these phenomena.

### 2.1 Structural and property evolution of $LiNiO_2$ cathode during operation

$LiNiO_2$ is a widely studied cathode material for lithium-ion batteries, undergoing complex structural transformations during charging and discharging processes, which substantially affect its electrochemical performance.[32-34] During charging, lithium ions are gradually extracted from the $LiNiO_2$ lattice, leading to the formation of $Li_{1-x}NiO_2$. With increasing lithium extraction (increasing x values), $LiNiO_2$ experiences a series of intricate phase transitions. For example, during charging, $LiNiO_2$ initially transforms from a hexagonal H1 phase into a monoclinic M phase, and subsequently transitions into hexagonal H2 and H3 phases.[35-36] Typically, undoped $LiNiO_2$ encounters multiple hexagonal and monoclinic phase regions throughout the charging process. Regarding lattice parameter evolution, $LiNiO_2$ undergoes significant changes along both the a-axis and c-axis between the H2 and H3 phases, with observed variations reaching approximately 7.6% along the a-axis and approximately 5.2% along the c-axis.[36] Simultaneously, $LiNiO_2$ may undergo irreversible structural alterations such as nickel-ion disordering and lattice oxygen loss during cycling.[37-40] Nickel-ion disordering refers to the migration of nickel ions from transition-metal layers into lithium layers, particularly under conditions of high-temperature synthesis or deep delithiation. Lattice oxygen loss occurs as a consequence of structural instabilities caused by complex phase transitions and the presence of highly oxidized nickel species (such as $Ni^{4+}$), resulting in oxygen release from the lattice.[37-40] Both nickel-ion disordering and lattice oxygen loss significantly affect lithium-ion transport: nickel-ion disorder blocks lithium-ion transport pathways, reducing electrochemical activity and cycle stability, whereas lattice oxygen loss creates oxygen vacancies that may cause structural collapse or undesired phase transformations.[37-40] These processes inhibit reversible lithium insertion and extraction, ultimately decreasing cycling stability and causing capacity loss. During discharge, lithium ions are re-inserted into the layered structure of $LiNiO_2$. Ideally, full lithium re-intercalation would restore the structure to a nearly pristine state. However, structural irreversibility induced during charging often prevents complete structural recovery, leading to gradual performance deterioration over extended cycling.[37] The voltage profile of $LiNiO_2$ during delithiation increases progressively, characterized by distinct voltage plateaus associated with specific phase transitions, such as the transformation from hexagonal to monoclinic phases.[41-42]

**2.2 Structural and property evolution of graphite anode during operation**

Graphite, a commonly employed anode material, also exhibits pronounced structural and property changes during lithium insertion and extraction.[43-50] During charging (lithiation), lithium ions progressively insert into graphite's interlayer spaces, forming a series of distinct lithium-graphite intercalation compounds (Li-GICs).[51-57] Continuous lithium insertion induces significant changes in graphite lattice parameters, especially along the c-axis, resulting in noticeable expansion of interlayer spacing. With increased lithium content, graphite forms multiple characteristic stage structures such as Stage I ($LiC_6$) and Stage II ($LiC_{12}$).[51-57] Each stage corresponds to distinct lithium concentrations and specific interlayer lithium arrangements.

During discharging (delithiation), lithium ions gradually leave the interlayer regions of graphite, causing a reduction in interlayer spacing, and lattice parameters revert to nearly their original dimensions. However, repeated cycling causes structural changes that are not fully reversible. Permanent defects or stacking faults may form within the graphite crystal structure due to the repeated expansion and contraction cycles.[58-60] This structural deterioration progressively reduces long-term cycling stability, manifesting as irreversible capacity loss.[61-62]

Lithium-ion transport properties within graphite also exhibit notable variation, strongly influenced by lithium-ion concentration between carbon layers. Specifically, the lithium-ion diffusion coefficient continuously decreases during lithiation and correspondingly increases during delithiation.[63] Additionally, due to the specific voltages associated with individual stage structures formed during cycling, the voltage profile of graphite exhibits distinct stepwise plateaus characteristic of each stage.[64]

**2.3 Computational simulation strategies for investigating structural and property evolution**

The structural transformations and electrochemical behavior of electrode materials during charge/discharge cycles involve intricate microscopic mechanisms, which directly govern their macroscopic performance. Computational simulation methods have rapidly developed into highly effective and reliable tools for investigating these intricate phenomena. Advanced computational techniques enable systematic simulations of lithium ion insertion and extraction within crystal structures, identifying optimal sites for lithium ion occupancy and elucidating the corresponding energy distributions and structural stability.

In addition to static structural considerations, the dynamical properties of lithium ion transport constitute another key area of computational investigation. Molecular dynamics (MD) simulations enable real-time tracking and quantitative assessment of lithium-ion migration behavior within electrode materials at various states of charge. Statistical analysis of mean squared displacement (MSD) data provides lithium-ion diffusion coefficients and reveals characteristic migration paths.[65-66] Such dynamic simulations clarify ion transport mechanisms and highlight how structural evolution and defect formation affect ionic mobility.

Moreover, computational methods can precisely quantify variations in lattice parameters of electrode materials under different states of charge, particularly distinguishing changes along specific crystallographic axes. Changes in lattice parameters closely relate to mechanical stress accumulation, structural expansion or contraction, and phase transitions. Detailed computational analysis of lattice parameter evolution clearly reveals internal structural dynamics, stress variations, and possible phase transitions. These analyses significantly enhance the understanding of structural stability and mechanical integrity during prolonged cycling.

Voltage profiles of electrode materials represent critical indicators of electrochemical performance, directly reflecting mechanisms of energy storage and release. Computational simulations systematically calculate the total energies of electrode materials at varying lithium concentrations, enabling accurate determination of theoretical voltage curves. Such theoretical calculations reliably capture phase transition plateaus, voltage hysteresis, and capacity trends throughout cycling, facilitating direct comparison with experimental results.

Additionally, computational simulations effectively elucidate lithium-ion distribution within electrode materials, including concentration gradients across different atomic layers or regions, and their evolution with state of charge. Statistical and data-mining approaches applied to structural data reveal lithium distribution characteristics and trends during cycling. Such analyses provide critical insight into ion migration tendencies, local lithium enrichment or depletion phenomena, and mechanisms underlying capacity fading and performance deterioration.

In this chapter, the structural and property evolution of lithium-ion battery electrode materials during charge/discharge cycles is systematically reviewed, with particular emphasis on $LiNiO_2$ as the cathode and graphite as the anode. The key mechanisms underlying phase transitions, lattice parameter changes, ion transport behavior, and voltage characteristics are elucidated, highlighting the direct influence of these physicochemical processes on electrochemical performance and cycling stability. Furthermore, this chapter introduces the application of advanced computational simulation methods for atomistic modeling of charge/discharge processes, investigation of lattice and voltage evolution, and exploration of ion transport and distribution. These discussions establish a solid theoretical foundation for the development and application of automated simulation platforms.

## 2 Module architecture

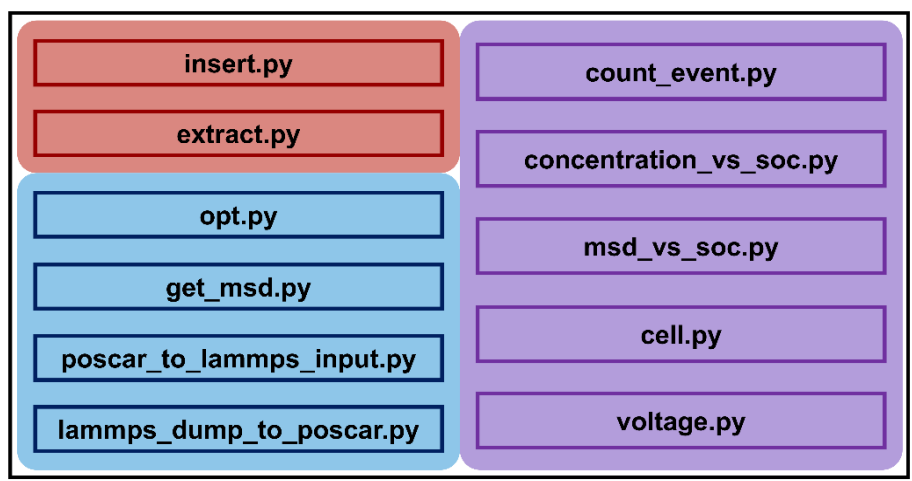

**Figure 1. Module Architecture of the EDIS Software. Red indicates the 2 core modules, blue represents the 4 auxiliary modules, and purple denotes the 5 property calculation and analysis modules.**

As research on lithium-ion batteries and their materials increasingly emphasizes multiscale, multi-physics coupling, and high-throughput analyses, the development of efficient and automated software tools has become critically important for achieving an in-depth understanding of material mechanisms and optimizing their performance.. This chapter systematically introduces the core architectural design and functional implementation of the EDIS software. Focusing on tasks such as lithium deintercalation simulations of electrode materials, property calculations, and analysis of property evolution, we elaborate on the implementation principles, data workflows, and applicability of 2 core modules and 5 property calculation and analysis modules, as well as describe the functions of 4 auxiliary modules (Figure 1).

### 3.1 Core module: lithium insertion module (insert.py)

The identification of optimal lithium insertion sites and the corresponding energy distribution within electrode materials is of great significance for enhancing the performance of lithium-ion batteries. To systematically analyze the lithium insertion process, we developed the insert.py module. This module is capable of automatically identifying all potential lithium insertion sites, predicting the total energy of each configuration following lithium insertion, and visualizing the resulting energy landscape in the form of a heatmap. Ultimately, it enables the determination of the most favorable lithium insertion site.

To enable efficient and comprehensive exploration of possible insertion positions, the module employs the grid search method as its core algorithm. The grid search method systematically identifies optimal insertion sites for target ions by uniformly distributing grid points throughout a three-dimensional spatial region. Specifically, the spatial boundaries and grid spacing are defined based on the unit-cell parameters and structural dimensions of the material, generating a three-dimensional grid covering the entire structural space. Subsequently, the distances between each grid point and the existing atoms are calculated, and points that are too close to atoms are excluded to avoid physical overlap and to improve computational efficiency. Users can flexibly adjust the grid spacing and filtering thresholds according to the specific characteristics of different materials and the required simulation accuracy, achieving an optimal balance between efficiency and precision. After filtering, the remaining valid grid points serve as candidate insertion sites. For each candidate site, a new structure is constructed and its total energy is calculated. The site with the lowest energy is ultimately identified as the optimal lithium insertion site.

```python
poscar_path = "POSCAR"
structure = read(poscar_path)
grid_spacing = 0.4
a, b, c = structure.cell.lengths()
num_p_a = 1
while True:
    if a/(num_p_a+1) >= grid_spacing:
        num_p_a += 1
    else:
        break
num_p_b = 1
while True:
    if b/(num_p_b+1) >= grid_spacing:
        num_p_b += 1
    else:
        break
num_p_c = 1
while True:
    if c/(num_p_c+1) >= grid_spacing:
        num_p_c += 1
    else:
        break
print(f"a, b, c directions require {num_p_a}, {num_p_b}, {num_p_c} grid points respectively")
```

In terms of implementation, the module first utilizes the ASE library to read the initial structure and extract the cell parameters and atomic coordinates. Based on the user-defined grid spacing, the number of grid points along the a, b, and c directions is determined, and a three-dimensional grid is generated to systematically sample the structural space. To avoid overlap with existing atomic positions, a distance-based threshold filtering strategy is applied, excluding redundant grid points within a specified cutoff distance (such as 1.5 Å). All parameters can be flexibly adjusted according to the specific material system.

```python
filtered_points = []
for point in grid_points:
    temp_structure = structure.copy()
    temp_structure += Atoms('H', positions=[point])
    distances = temp_structure.get_distances(-1, range(len(structure)), mic=True)
    min_distance = np.min(distances)
    if min_distance > 1.5:
        filtered_points.append(point)
    temp_structure.pop()
```

After filtering, a temporary structure is generated for each valid grid point, and, in conjunction with machine learning potentials, the total energies of all candidate lithium-inserted structures are batch-calculated. All energy and coordinate data are exported to a text file for subsequent analysis. The program automatically identifies the lowest-energy site, outputs the corresponding structure in CONTCAR format, and records all grid points and their energies for traceability.

```python
def calculate_energy(point):
    temp_structure = structure.copy()
    temp_structure += Atoms('Li', positions=[point])
    temp_poscar_path = 'CONTCAR'
    write(temp_poscar_path, temp_structure, format='vasp')
    system = dpdata.System(temp_poscar_path, fmt='poscar')
    labeled_system = system.predict(dp_model)
    energy = float(labeled_system['energies'][0])
    return point, energy
energies = []
for point in filtered_points:
    try:
        result = calculate_energy(point)
        energies.append(result)
    except Exception as e:
        print(f"Error calculating energy for point {point}: {e}")
if not energies:
    print("Error: No energies calculated. Check the DeepPot model and input structure.")
    exit()
min_energy_point, min_energy = min(energies, key=lambda x: x[1])
best_structure = structure.copy() + Atoms('Li', positions=[min_energy_point])
```

Finally, the module employs the Plotly library to generate an interactive three-dimensional energy heatmap, which intuitively visualizes the spatial energy distribution of lithium ions at different positions. The heatmap is saved in HTML format, allowing users to explore the data interactively within a web browser, supporting multi-angle rotation, zooming, and filtering. This provides valuable visual support for the structural optimization and performance improvement of electrode materials.

```python
x_coords = [point[0] for point, _ in energies]
y_coords = [point[1] for point, _ in energies]
z_coords = [point[2] for point, _ in energies]
energy_values = [energy for _, energy in energies]
fig = go.Figure()
```

**3.2 Core module: lithium extraction module (extract.py)**

To facilitate the investigation of lithium-ion extraction mechanisms in electrode materials during charge/discharge processes, we have developed the extract_lithium.py module. This module is designed to systematically predict and screen all possible lithium extraction sites, ultimately determining the optimal extraction site. In doing so, it provides a powerful tool for studying lithium migration mechanisms and assessing the structural stability of electrode materials.

The module employs the queue-bubbling method, which is a systematic approach for structural energy screening and ranking, specifically developed for identifying the most favorable extraction sites for target ions such as lithium. In practical applications, for example with lithium ion battery cathode materials, the method operates by sequentially removing each lithium atom from the structure to generate a batch of candidate one-lithium-extracted configurations. The total energy of each candidate is then predicted using efficient energy evaluation tools such as machine learning interatomic potentials. The energies of all configurations are automatically recorded and sorted. Through iterative bubbling, the configuration with the lowest energy is progressively moved to the top of the queue, thus identifying the optimal extraction site and its corresponding structure. This approach enables a comprehensive traversal and comparison of all possible extraction pathways, significantly enhancing both the efficiency and accuracy of identifying the energetically most favorable site.

```python
for li_index in li_indices:
    modified_structure = structure.copy()
    modified_structure.remove_sites([li_index])
    temp_poscar_path = "CONTCAR"
    Poscar(modified_structure).write_file(temp_poscar_path)
    system = dpdata.System(temp_poscar_path)
    labeled_sys_yes_500w = system.predict(dp_yes_500w)
    energy = float(labeled_sys_yes_500w['energies'][0])
    energies.append(energy)
    structures.append(modified_structure)
    os.remove(temp_poscar_path)
min_energy_index = energies.index(min(energies))
optimal_structure = structures[min_energy_index]
optimal_li_index = min_energy_index + 1
print(f"best Li index: {optimal_li_index}")
print(f"lowest energy number: {optimal_structure}")
optimal_poscar = Poscar(optimal_structure)
optimal_poscar.write_file("CONTCAR")
```

In terms of implementation, the module first reads the user-supplied structure file in POSCAR format and utilizes the Pymatgen library to parse the structure and extract the indices of all lithium atoms. After loading the machine learning potential, the program iteratively processes each lithium atom, generating a new delithiated structure for each possible extraction scenario, thereby systematically covering all candidate configurations. For every generated structure, the program invokes the trained interatomic potential to predict the total energy, automatically recording the energy of each configuration. Upon completion of the global comparison, the structure with the lowest energy is identified as the optimal delithiated configuration, and the corresponding atomic site is recognized as the most favorable extraction position.

```python
with open("energies.txt", "w") as f:
    for i, (index, energy) in enumerate(zip(li_indices, energies)):
        li_number = i + 1
        if i == min_energy_index:
            f.write(f"Li site {li_number}: {energy} (minimum energy)\n")
        else:
            f.write(f"Li site {li_number}: {energy}\n")
file_contcar = open("CONTCAR")
file_content = file_contcar.read()
with open("CONTCAR", "w") as f:
    f.write(file_content.replace(" Li", ""))
print("All energy has saved to energies.txt")
```

Finally, the optimal delithiated structure is output in VASP CONTCAR format to facilitate subsequent structural analysis and kinetic simulations. The module also automatically generates an energy record file that summarizes the energies of all candidate structures and highlights the site with the lowest energy, enabling users to quickly locate key data. Researchers can flexibly utilize this script for different electrode materials and charge/discharge states to perform automated simulations of lithium extraction processes according to their specific research needs.

### 3.3 Property analysis and calculation modules

To enable comprehensive analysis of structural and property evolution in electrode materials during lithium insertion and extraction, the EDIS platform integrates a series of auxiliary analysis modules. These modules span atomic layer identification, lithium distribution statistics, cell parameter analysis, and voltage curve calculations, thereby enriching the data mining and visualization capabilities of high-throughput dynamic simulations.

#### *3.3.1 Lithium insertion/extraction position statistics and atomic layer identification module (count_event.py)*

This module is designed for layered systems and utilizes the K-means clustering algorithm to automatically identify atomic layers and count the number of Li ions between each pair of adjacent layers.[68-69] This enables spatial distribution analysis of lithium ions during the charge/discharge process. Users can customize the criteria for layer identification and statistical analysis according to the material structure, with support for probabilistic normalization to quantitatively reflect the dynamic frequency of lithium insertion and extraction events at different positions. The module is particularly suitable for systems with well-defined atomic layers. For non-ideal layered materials, the clustering and statistical logic can be adjusted to accommodate actual structural features.

```python
li_positions = [site.frac_coords[2] for site in structure if site.species_string == "Li"]
c_positions = [site.frac_coords[2] for site in structure if site.species_string == "C"]
c_positions = sorted(c_positions)
num_layers = 8
kmeans = KMeans(n_clusters=num_layers, random_state=0)
c_positions = [[pos] for pos in c_positions]
kmeans.fit(c_positions)
labels = kmeans.labels_
```

#### *3.3.2 Lithium layer concentration and SOC correlation analysis module (concentration_vs_soc.py)*

```python
sorted_layers = sorted(layer_dict.items(), key=lambda x: sum(x[1]) / len(x[1]))
for i, (layer_index, positions) in enumerate(sorted_layers):
    average_c = sum(positions) / len(positions)
    average_c_list.append(average_c)
    if len(positions) != 54:
        fault_number += 1
for li_position in li_positions:
    if (li_position > average_c_list[0]) and (li_position < average_c_list[1]):
        layer_2_li += 1
    elif (li_position > average_c_list[1]) and (li_position < average_c_list[2]):
        layer_3_li += 1
    else:
        layer_5_li += 1
layer_1.append(100*layer_1_li/9)
layer_2.append(100*layer_2_li/9)
layer_3.append(100*layer_3_li/9)
layer_4.append(100*layer_4_li/9)
```

To dynamically track the evolution of lithium layers during the charge/discharge cycle, this module integrates K-means clustering and statistical analysis to automatically quantify and visualize the concentration of lithium in each atomic layer under different states of charge (SOC).[70] The workflow includes structural clustering, stepwise counting of Li in each layer, and visualization of the evolutionary trends using Matplotlib.[71] With flexible parameter settings, the module is applicable to ordered layered structures such as graphite and LiNiO$_2$, revealing interlayer migration and distribution features and providing data support for analyzing the relationship between lithium extraction mechanisms and SOC.

### *3.3.3 Elemental mean squared displacement and SOC correlation analysis module (msd_vs_soc.py)*

```python
def load_and_process_file(file_path):
    data = pd.read_csv(
        file_path,
        comment='#',
        delim_whitespace=True,
        names=['TimeStep', 'c_msd1[4]', 'c_msd2[4]']
    )
    data = data[data['TimeStep'] > 5000].reset_index(drop=True)
    data['TimeStep'] = data['TimeStep'] - data['TimeStep'].iloc[0]
    return data

def combine_structures(file_paths):
    combined_data = pd.DataFrame()
    segment_boundaries = []
    for i, file_path in enumerate(file_paths):
        data = load_and_process_file(file_path)
        data['TimeStep'] += i * 30000
        data['TimeStep'] = data['TimeStep'] / 3240000 * 100
        if i > 0:
            segment_boundaries.append(data['TimeStep'].iloc[0])
        combined_data = pd.concat([combined_data, data], ignore_index=True)
    return combined_data, segment_boundaries
```

This module is specifically designed to elucidate the rules governing ion migration and structural response in electrode materials. It automatically reads and integrates MSD (mean squared displacement) data across different SOC stages, enabling multi-element, multi-cycle trend analysis and visualization. The module supports batch file processing, data normalization, and automatic segmented plotting to ensure consistent comparison of data from different stages within a unified coordinate system. For example, in LiNiO$_2$, it can generate MSD evolution curves for Li, Ni, and other elements, supports the customization of demarcation lines and multiple plotting parameters, and outputs publication-quality analytical results. It is suitable for performance evaluation and kinetic mechanism studies.

### 3.3.4 Voltage curve calculation and analysis module (voltage.py)

Voltage variation in electrode materials is a core indicator for evaluating energy storage mechanisms and performance. This module automatically reads a series of structural and energy data, performs energy normalization for different lithium contents (e.g., x in Li$_x$NiO$_2$), calculates segmented voltage values, and plots voltage-composition curves. Using a segmental difference method, it calculates the voltage for each (dis)charge interval, clearly depicting charge/discharge plateaus, phase transition regions, and voltage hysteresis.[72] All input paths, normalization, and plotting parameters are fully customizable, making the module adaptable to various systems and providing a crucial data foundation for lithium storage mechanism analysis and experimental comparison.

```python
energies_per_fu_ex = []
li_x_list_ex = []
for idx, file_path in enumerate(files_ex):
    structure = Structure.from_file(file_path)
    n_atoms = len(structure)
    n_Li = sum([site.species_string == "Li" for site in structure])
    n_Ni = structure.composition["Ni"]
    x = n_Li / n_Ni
    li_x_list_ex.append(x)
    with open(r"energies.txt") as f:
        for i, line in enumerate(f):
            if i == idx:
                eV_per_atom = float(line.strip())
                break
    e_total = eV_per_atom * n_atoms
    e_per_fu = e_total / n_Ni
    energies_per_fu_ex.append(e_per_fu)
E_li_metal = -1.879694
voltage_ex = []
x_axis_ex = []
for i in range(1, len(files_in)):
    x1, x2 = li_x_list_ex[i-1], li_x_list_ex[i]
    E1, E2 = energies_per_fu_ex[i-1], energies_per_fu_ex[i]
    delta_x = x2 - x1
    if abs(delta_x) < 1e-6:
        voltage_ex.append(None)
        x_axis_ex.append((x1 + x2) / 2)
        continue
    v = - (E2 - E1 - (x2 - x1) * E_li_metal) / delta_x
    voltage_ex.append(v)
    x_axis_ex.append((x1 + x2) / 2)
```

### 3.3.5 Cell parameter evolution analysis module (cell.py)

To quantitatively analyze changes in lattice constants during the charge/discharge process, the cell.py module enables automatic extraction, statistical analysis, and visualization of cell parameters from structural files at different SOCs. It can batch-match SOC and structural data, summarize and plot the evolution trends of the a, b, and c lattice parameters, and specifically highlight the variation of the c-axis with SOC. With flexible configuration of all input and plotting parameters, the module is suitable for a wide range of material systems, provides direct evidence for structural evolution, phase transitions, and anisotropic responses, and achieves publication-quality visualization.

```python
files_ex = []
for i in range(your_number):
    files_ex.append(f"your_path/{i}/CONTCAR")
a_list_ex = []
b_list_ex = []
c_list_ex = []
for file_path in files_ex:
    try:
        atoms = read(file_path, format='vasp')
        cell = atoms.cell.lengths()
        a_list_ex.append(cell[0])
        b_list_ex.append(cell[1])
        c_list_ex.append(cell[2])
    except Exception as e:
        print(f"Error reading {file_path}: {e}")
        a_list_ex.append(None)
        b_list_ex.append(None)
        c_list_ex.append(None)
```

**3.4 Auxiliary modules**

To further enhance the efficiency and automation of electrode material simulation and analysis, this work has also developed a series of auxiliary modules, covering functionalities such as structure optimization, dynamic analysis, format conversion, and lithium extraction analysis. These modules not only enable highly efficient coordination of data processing workflows, but also provide robust tools for the analysis of material structures and dynamic properties across multiple computational platforms.

*3.4.1 Structure optimization module (opt.py)*

```python
def fix_cell_angles(unit_cell):
    cell_params = cell_to_cellpar(unit_cell)
    fixed_angles = cell_params[3:]
    def set_variable_lengths(new_lengths):
        return cellpar_to_cell([new_lengths[0], new_lengths[1], new_lengths[2], *fixed_angles])
    return set_variable_lengths
```

This module enables efficient batch structural optimization of electrode materials. It is based on machine learning potentials (such as CHGNet or DeepMD) and the ASE library, and employs the BFGS optimization algorithm to simultaneously optimize both cell parameters and atomic coordinates.[73-74] By implementing a custom fix_cell_angles function, the module ensures that cell angles remain unchanged during the optimization process, effectively preventing structural distortion and improving the stability and reliability of the optimization. The module supports automatic recognition and batch processing of structural files, with all optimized results output in the unified CONTCAR format, greatly streamlining the preprocessing workflow for complex systems.

```python
def batch_optimize_structure(file_paths, model_path, type_map, max_force=0.02):
    calculator = DP(model=model_path, type_map=type_map)
    for file_path in file_paths:
        optimize_structure(file_path, calculator, max_force)

def optimize_structure(file_path, calculator, max_force=0.02):
    print(f"Start optimizing: {file_path}")
    structure = read(file_path)
    structure.set_calculator(calculator)
    ucf = UnitCellFilter(structure)
    optimizer = BFGS(ucf, trajectory='optimization.traj')
    optimizer.run(fmax=max_force)
    optimized_path = file_path.replace("POSCAR", "CONTCAR")
    write(optimized_path, structure)
    print(f"Optimization complete, results saved at: {optimized_path}")
```

*3.4.2 Mean squared displacement analysis module (get_msd.py)*

```python
paths = [r"\X\out.msd", r"\XX\out.msd", r"\XXX\out.msd"]
for i, path in enumerate(paths):
    data = np.loadtxt(path, skiprows=2)
    time = data[:, 0] / 1000
    msd1 = data[:, 1]
    msd2 = data[:, 2]
    plt.plot(time, msd1, linestyle='-', color=colors[i], linewidth=3, label=f'XXX Li')
    plt.plot(time, msd2, linestyle='--', color=colors[i], linewidth=3, label=f'XXX C')
```

This module is specifically designed for the automated statistical analysis and visualization of MSD data obtained from molecular dynamics simulations. It can batch read out.msd files from various systems, automatically separate the MSD information for different elements, and generate high-quality plots using Matplotlib. To ensure clarity and aesthetics, the module standardizes axis labels, fonts, tick marks, and legends, with all analytical results exported in high-resolution PNG format. This provides a solid data foundation for subsequent analyses of ion migration behavior.

*3.4.3 Structure format conversion module (poscar_to_lammps_input.py)*

This module enables efficient interconversion between VASP and LAMMPS structure files. It automatically identifies atomic species, spatial coordinates, and periodic boundary conditions in POSCAR files, and reconstructs structures and exports data via ASE objects. The module supports automatic correction of atomic types and related parameters, ensuring that the generated LAMMPS input files can be directly used in molecular dynamics simulations, thereby significantly enhancing the automation of multi-platform collaborative research.

```python
def poscar_to_lammps(poscar_file, lammps_file):
    atoms = read(poscar_file)
    new_atoms = Atoms(
        symbols=atoms.get_chemical_symbols(),
        positions=atoms.get_positions(),
        cell=atoms.get_cell(),
        pbc=atoms.get_pbc()
    )
    write(lammps_file, new_atoms, format='lammps-data')
```

### *3.4.4 Trajectory format conversion and element correction module (lammps_dump_to_poscar.py)*

```python
input_dump_file = "out.dump"
frames = read(input_dump_file, index=':')
for i, frame in enumerate(frames):
    os.makedirs(str(i))
    write(str(i)+"/POSCAR", frame, format='vasp')
    print(f'Frame {i} written to {str(i)+"/POSCAR"}')
    file_poscar = open(str(i)+"/POSCAR")
    file_content = file_poscar.read()
    with open(str(i)+"/POSCAR", "w") as file_poscar:
        file_poscar.write(file_content.replace("H ", "Li ").replace("He ", "C "))
```

This module supports batch conversion of LAMMPS trajectory outputs (out.dump) to VASP POSCAR format and automatically corrects element types. To address issues where ASE misidentifies atomic types when parsing trajectories (e.g., recognizing type 1 and 2 as H and He), the module offers a flexible element name replacement scheme, allowing users to adjust rules according to specific systems and ensuring the accuracy of multi-element structural files. All structural files are output frame-by-frame in separate folders, facilitating seamless integration with subsequent analysis and simulation workflows.

In this chapter, we systematically describe the modular architecture of the EDIS software and the functionalities of its constituent modules, with particular emphasis on its design for efficient and automated simulation of lithium-ion battery electrode materials. EDIS integrates two core modules (lithium insertion and extraction), four property calculation and analysis modules, and four auxiliary modules encompassing structure optimization, data processing, and format conversion. Through the coordinated operation of these modules, the platform enables automated identification and energy evaluation of lithium (de)intercalation sites, comprehensive analysis of structural and electrochemical property evolution, and high-throughput data mining across various material systems. By integrating machine learning interatomic potentials and advanced data visualization tools, EDIS supports rigorous, reproducible, and extensible research workflows for investigating ion migration mechanisms, phase transitions, and performance-related properties. The unified and extensible system architecture of EDIS not only significantly enhances the efficiency of electrode material simulations, but also provides a robust foundation for systematic studies of structure–property relationships.

## 4. Workflow and application examples

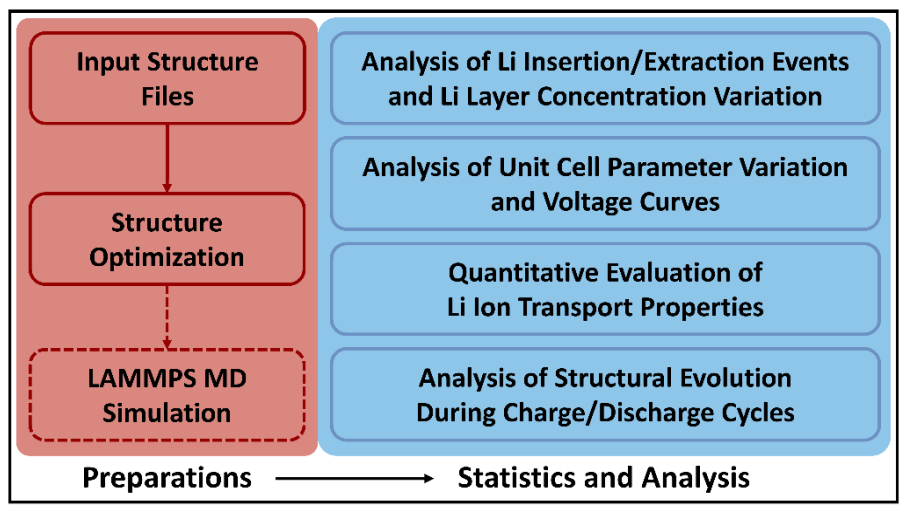

Figure 2: Workflow and Functional Overview of EDIS.

The overall workflow and functional architecture of the EDIS software exemplify its integrated design philosophy for the simulation and mechanistic analysis of electrode material dynamics. As illustrated in Figure 2, the entire process begins with the structural optimization of the input material. Users can utilize machine learning potentials (such as CHGNet) to perform variable-cell structural optimization on the selected electrode material system, thereby obtaining a more stable initial configuration that more closely reflects the actual physical state. This step provides a solid foundation for subsequent molecular dynamics simulations and various physicochemical analyses. Upon completion of structural optimization, users may choose either to invoke the LAMMPS platform for molecular dynamics simulations or to directly conduct further analyses based on the optimized structure. When employing LAMMPS for molecular dynamics simulations, parameters such as temperature, pressure, simulation timestep, and sampling intervals can be flexibly adjusted according to specific research needs, thus enabling the investigation of ion transport behaviors in various material systems under different conditions. The structural evolution data collected from molecular dynamics simulations or structural optimization serves as the basis for subsequent multidimensional analyses.

Following the preparatory steps, EDIS offers a series of optional advanced functions that enable users to conduct targeted analyses tailored to specific scientific questions. The software features four main analytical capabilities, grouped according to their data sources and application scenarios. First, EDIS supports analysis of lithium insertion/extraction events and lithium layer concentration variation, utilizing either LAMMPS-generated trajectory files or intermediate structures obtained from structural optimization. For lithium insertion, the software generates potential energy surface heatmaps and provides detailed records of insertion site distributions, while for lithium extraction, it outputs the corresponding atomic indices and frequency distributions of extraction events. Second, EDIS enables the analysis of unit cell size variations and voltage curve calculations based on structures derived from either LAMMPS molecular dynamics simulations or structural optimization. These first two analytical functions are compatible with both types of structural data. In contrast, the third and fourth functions are available exclusively for results obtained through LAMMPS molecular dynamics simulations. Specifically, EDIS allows for the quantitative assessment of lithium-ion transport properties by automatically calculating the MSD of each element under user-defined simulation conditions, thereby reflecting the diffusion behavior of different species during the simulation. The resulting trajectory files can be further visualized in three dimensions using specialized software such as OVITO, facilitating microscopic mechanistic analysis. Additionally, EDIS provides comprehensive quantitative analysis of structural evolution throughout the charge and discharge cycles, offering trend curves for the MSD of each element, the number (or concentration) of lithium ions within each atomic layer, and the system's total energy as a function of SOC. By analyzing these evolving datasets, users can gain in-depth insights into the coupling between microstructural responses and ion transport properties at various steps of the charge/discharge process.

To further demonstrate the practical value of EDIS and enhance the clarity and reproducibility of its workflow, the following sections will use $LiNiO_2$ cathode materials as an example to provide a detailed description of typical usage strategies and analytical approaches for each functional module. This representative case study is intended to serve as a reference for researchers in related fields and to promote the broader application and adoption of EDIS across various electrode material systems.

## 4.1 Structural optimization

In studies of the dynamic mechanisms of LiNiO$_2$ cathode materials, high-quality structural input and a well-designed simulation workflow are essential prerequisites for ensuring the reliability of analytical results. First, variable-cell structural optimization is performed on the LiNiO$_2$ supercell structure (Li$_{27}$Ni$_{27}$O$_{54}$) using machine learning potentials. This process uses a POSCAR file in VASP format as input and calls the opt.py module in the EDIS software together with the pre-trained CHGNet machine learning potential to predict and optimize the structural energies and forces.[75] Upon completion of the optimization, the resulting CONTCAR file represents a new, stable structure that more accurately reflects the atomic arrangement and crystal parameters of the material.

## 4.2 Simulation of charge/discharge process

During the charge/discharge cycles, the internal structure of the LiNiO$_2$ cathode undergoes significant changes as lithium ions are inserted and extracted. EDIS provides a fully automated analysis platform for this complex process, enabling users to systematically reveal the synergistic evolution of structural changes and transport properties by integrating batch scripts, dynamic simulations, and multidimensional data analysis.

```bash
for i in $(seq 0 27); do
    mkdir -p $i
    cd $i
    cp ../POSCAR_transfer POSCAR
    cp ../opt.py .
    cp ../extract.py .
    python3 opt.py
    python3 extract.py
    cp CONTCAR_extracted ../POSCAR_transfer
    cd ..
done
```

To simulate the continuous extraction of lithium, we developed a .sh script as shown above. The script uses the optimized Li$_{27}$Ni$_{27}$O$_{54}$ supercell structure as the initial state and sequentially calls the extract.py module to remove lithium, followed by the opt.py module to optimize the structure after each extraction. This cycle is repeated until all lithium atoms are extracted from the structure. All intermediate and final structures generated during this process are saved for subsequent analysis.

Representative illustrations of the initial structure, some intermediate structures, and the final structure are shown in Figure 3.

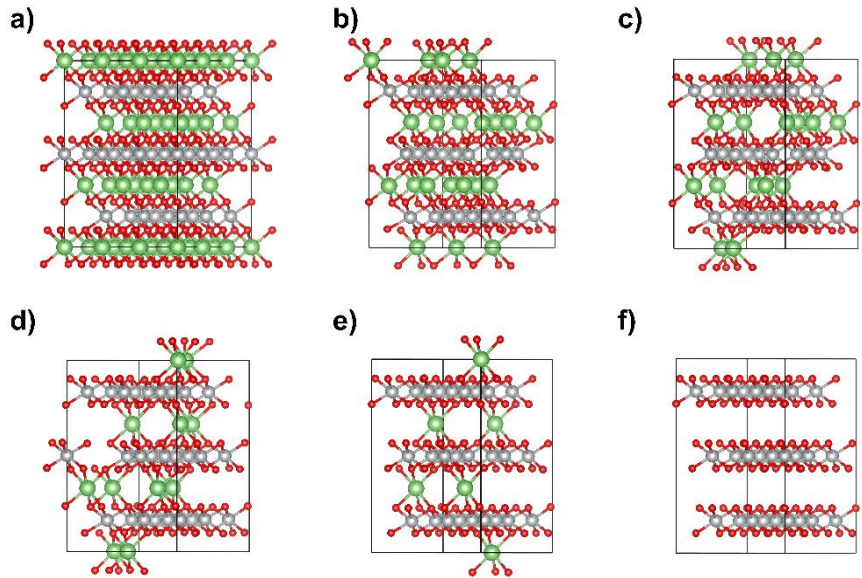

Figure 3: Sequential structural evolution of LiNiO$_2$ during continuous lithium extraction: (a) Li$_{27}$Ni$_{27}$O$_{54}$, (b) Li$_{21}$Ni$_{27}$O$_{54}$, (c) Li$_{16}$Ni$_{27}$O$_{54}$, (d) Li$_{11}$Ni$_{27}$O$_{54}$, (e) Li$_6$Ni$_{27}$O$_{54}$, and (f) Ni$_{27}$O$_{54}$.

```
for i in $(seq 0 27); do
    mkdir -p $i
    cd $i
    cp ../POSCAR_transfer POSCAR
    cp ../opt.py .
    cp ../insert.py .
    python3 opt.py
    python3 insert.py
    cp CONTCAR_inserted ../POSCAR_transfer
    cd ..
done
```

Similarly, the simulation of the continuous lithium insertion process is also implemented using a .sh script, as shown above. The optimized lithium-extracted end structure, Ni$_{27}$O$_{54}$, serves as the initial state for lithium insertion. The script sequentially calls the insert.py and opt.py modules to perform lithium insertion and subsequent structural optimization, respectively. This cycle is repeated until lithium insertion reaches saturation (Li$_{27}$Ni$_{27}$O$_{54}$). A series of intermediate and final structures are saved throughout the process for subsequent analysis. Representative illustrations of the initial structure, some intermediate structures, and the final structure are shown in Figure 4.

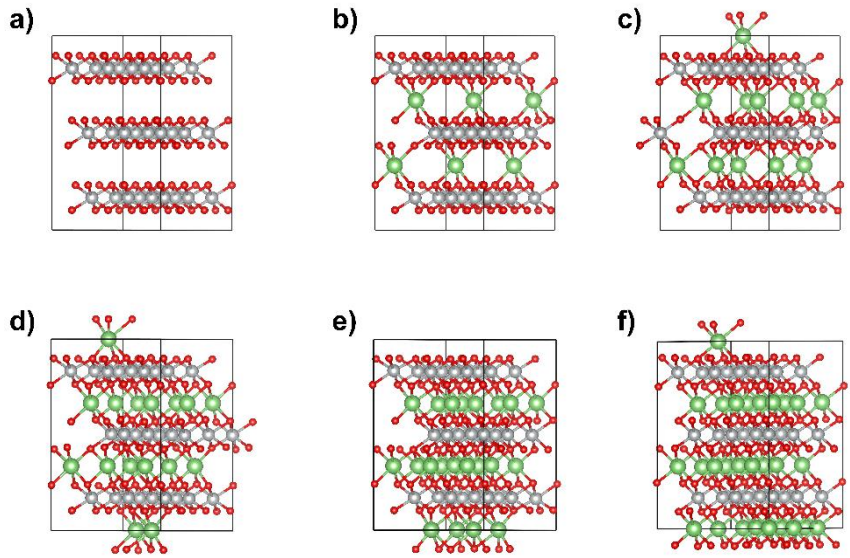

**Figure4:** Sequential structural evolution of LiNiO$_2$ during continuous lithium insertion: (a) Ni$_{27}$O$_{54}$, (b) Li$_6$Ni$_{27}$O$_{54}$, (c) Li$_{11}$Ni$_{27}$O$_{54}$, (d) Li$_{16}$Ni$_{27}$O$_{54}$, (e) Li$_{21}$Ni$_{27}$O$_{54}$, and (f) Li$_{27}$Ni$_{27}$O$_{54}$.

## 4.3 Analysis of lithium layer concentration changes during charge/discharge cycle

Analyzing the concentration changes of lithium within each layer of the structure can provide valuable insights into lithium ion migration pathways, phase transition behaviors, and structural stability in electrode materials. The EDIS platform enables efficient and automated analysis of lithium layer concentration variations throughout the charge/discharge cycles. Once a series of structural evolution data for LiNiO$_2$ cathode materials during cycling has been collected, the concentration_vs_soc.py script can be used to calculate and plot the concentration changes of each lithium atomic layer, thereby providing a comprehensive visualization of the temporal evolution of lithium distribution within the structure.

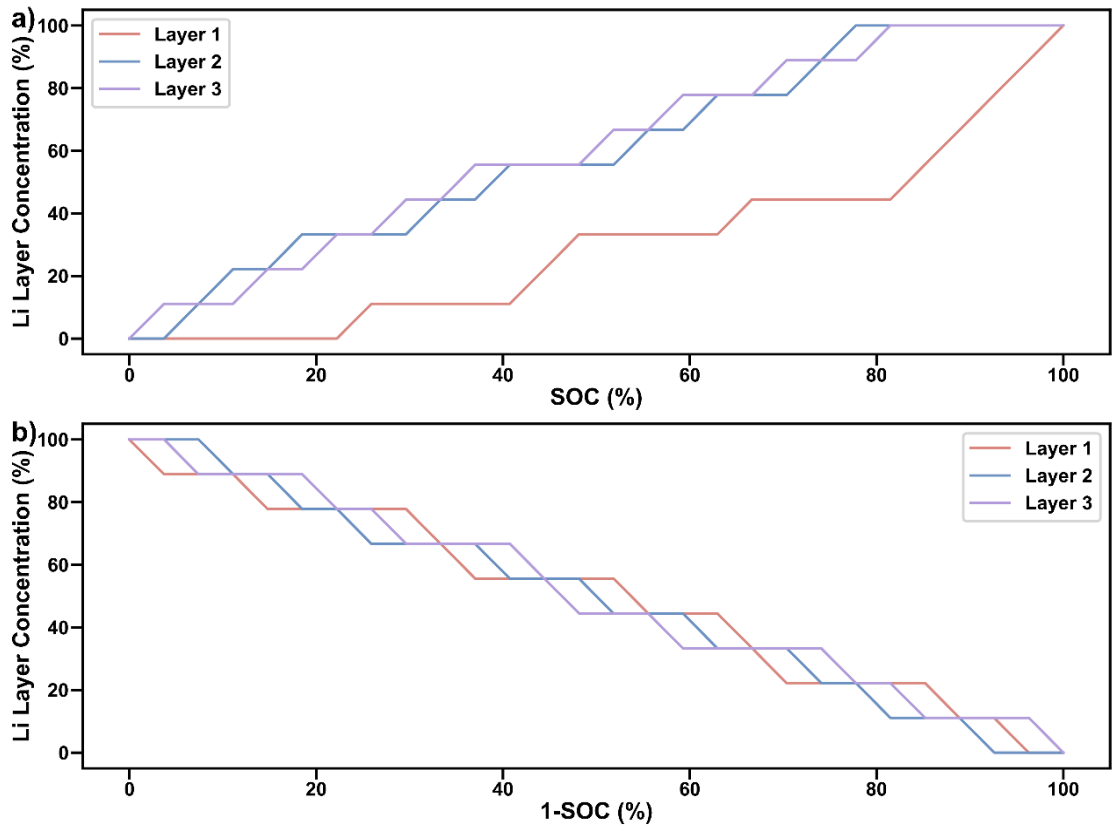

**Figure 5:** Layer-resolved lithium concentration profiles in LiNiO$_2$ during (a) lithium extraction and (b) lithium insertion, illustrating the temporal evolution and spatial distribution of lithium ions across different layers.

As shown in Figure 5a, for the lithium extraction process in LiNiO$_2$, the results indicate that the extraction probabilities for lithium ions in each layer are approximately the same, resulting in a uniform decrease in the concentration of all layers. However, as shown in Figure 5b, during the lithium insertion process, the concentrations of the individual layers do not increase uniformly; rather, the concentrations of Layer 2 and Layer 3 rise more rapidly and reach saturation first. This phenomenon reflects the spatial inhomogeneity present during lithium insertion.

### 4.4 Analysis of unit cell size variation

Analyzing changes in unit cell dimensions can provide useful insights into the volumetric response, structural stability, and phase transition mechanisms of materials during lithium insertion and extraction. The EDIS platform supports automated analysis of changes in unit cell parameters along different crystallographic directions throughout the charge/discharge cycles. Once a series of structural evolution data for LiNiO$_2$ cathode materials during cycling has been collected, the cell_vs_soc.py script can be used to extract and plot the variation in unit cell dimensions as a function of SOC, thereby clarifying the effects of lithium insertion and extraction on the structure.

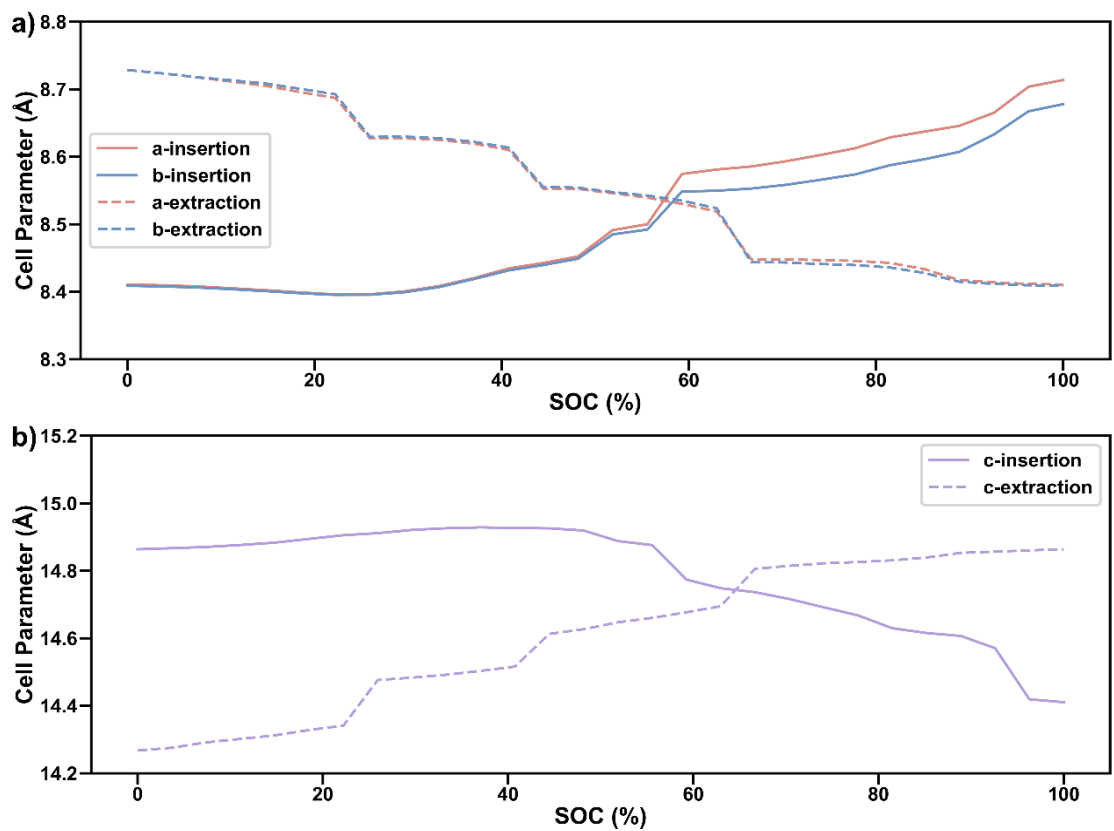

**Figure 6: Variation of unit cell dimensions in LiNiO$_2$ during lithium extraction and insertion processes: (a) a and b axes, (b) c axis.**

As shown in Figure 6a, due to the continuous variation of intralayer repulsive forces associated with lithium insertion and extraction, the unit cell dimensions parallel to the atomic layers (a and b directions) also undergo continuous and monotonic changes. However, in the direction perpendicular to the atomic layers (c direction, Figure 6b), a non-monotonic variation is observed during lithium insertion, indicating that the c axis exhibits a more complex response to lithium insertion and extraction.

**4.5 Voltage Curve Calculation**

Analyzing the voltage profile of electrode materials is crucial for understanding their electrochemical performance and practical applicability in lithium-ion batteries. The EDIS platform supports automated calculation and visualization of voltage curves throughout the charge/discharge cycles. Once a series of structural evolution data for

LiNiO₂ cathode materials during cycling has been collected, the voltage.py script can be used to calculate the voltage based on the relationship between energy and voltage and to plot the voltage profile as a function of the lithium molar fraction.

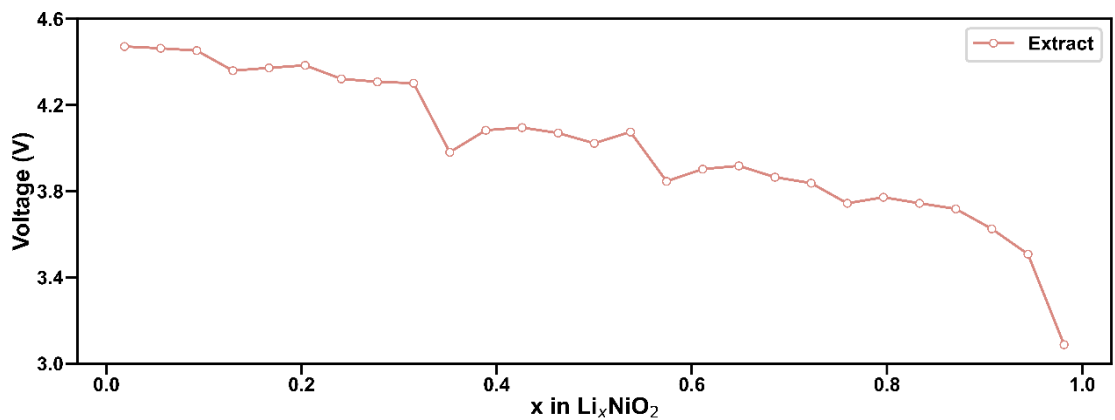

**Figure 7: Calculated voltage profile for LiNiO₂ during the lithium extraction process.**

As shown in Figure 7, the voltage curve for the lithium extraction process closely matches the experimentally measured voltage range. [76]

In this chapter, we systematically introduced the overall workflow, core functionalities, and practical application examples of the EDIS software for electrode material dynamics simulation and mechanistic analysis. EDIS integrates machine learning-based structural optimization, molecular dynamics simulations, and automated multidimensional data analysis to support comprehensive investigations of lithium-ion migration, structural evolution, and electrochemical properties in electrode materials. Using LiNiO₂ cathode materials as a representative case, we demonstrated the end-to-end process from structure preparation and charge/discharge simulation to advanced analyses such as lithium layer concentration evolution, unit cell parameter variation, and voltage curve calculation. These workflows not only improve the efficiency and reproducibility of data-driven research but also provide theoretical guidance and practical tools for elucidating complex physicochemical phenomena in rechargeable battery materials. Collectively, the application examples presented in this chapter highlight the versatility, scalability, and scientific value of the EDIS platform, thereby promoting its broader adoption in the field of energy materials research.

5. Conclusion and outlook

This work presents the systematic development and introduction of EDIS, the Electrode Dynamic Ion Intercalation/Deintercalation Simulator, to address the multiscale kinetic simulation requirements for ion insertion and extraction in electrode materials. EDIS integrates structural optimization, molecular dynamics simulation, data processing, and multidimensional analysis, providing an efficient and automated research tool for studying the dynamic behaviors and structural evolution of lithium-ion battery electrode materials during charge and discharge cycles.

In the theoretical foundations section, we used representative systems such as $LiNiO_2$ cathodes and graphite anodes as examples to illustrate the structural and property changes occurring during cycling. By examining these typical systems, we clarified the key scientific questions and achievable goals for computational simulation, such as phase transitions, changes in lattice parameters, ion transport, and voltage characteristics. This approach provided clear guidance for the functional design and analytical workflow of EDIS, ensuring that the software effectively supports mechanistic studies and structure–property relationship analysis.

In the module architecture section, we systematically introduced the core design concepts and implementation logic of EDIS. The software integrates modules for automatic screening of lithium insertion and extraction sites, structural and property analysis, and auxiliary functions such as structure optimization and format conversion. Its modular and extensible architecture guarantees adaptability and flexibility, making the platform suitable for a wide range of material systems and research needs.

In the workflow and application examples section, we demonstrated the complete EDIS workflow using $LiNiO_2$ as a case study. This included structure preparation, parameter setup, automated simulation, and multidimensional data analysis. Through systematic analysis of lithium layer concentration, cell parameter evolution, and voltage curves, we provided a detailed understanding of the coupling between ion migration and structural responses. We also employed EDIS to simulate the lithiation/delithiation process in graphite, achieving insights into the stage transformation mechanism.[77] The automated workflows and statistical tools significantly improved simulation efficiency and reproducibility, enabling rapid screening and quantitative evaluation in high-throughput and multi-parameter systems.

Although EDIS has already demonstrated strong versatility and automation, there is still room for improvement. Future developments may include the integration of multiscale simulation algorithms to overcome limitations in system size and simulation timescale. There is also a need to develop more general algorithms for event recognition and layer identification in complex multiphase or interfacial systems. In addition, combining big data visualization and AI-assisted analysis will further enhance the intelligence and applicability of the platform.

In a word, EDIS and its automated analysis workflows provide an efficient and extensible solution for multiscale kinetic simulation and data-driven research on electrode materials. As the platform continues to expand its capabilities, EDIS is expected to play an increasingly important role in energy materials, energy storage devices, and interfacial science, driving the integration of theoretical innovation and engineering practice in materials research.

**Acknowledgements**

This work was financially supported by the Strategic Priority Research Program of Chinese Academy of Sciences (grant no. XDB1040300), and the National Natural Science Foundation of China (grants no. 52172258). We acknowledge the National Supercomputer Center in Tianjin and Bohrium AI for Science Research Space Station for providing computational resources.